\documentclass[preprint,12pt]{elsarticle}

\usepackage{amssymb}
\usepackage{amsmath}

\usepackage{lineno}

\journal{Journal of Subatomic Particles and Cosmology}

\begin{document}

\begin{frontmatter}



\title{Precise measurement of the $t\bar{t}$ production	cross-section and lepton differential	distributions in $e\mu$ dilepton events}


\author{Dominic Hirschb\"uhl} 
\author{on behalf of the ATLAS collaboration}
\affiliation{organization={University of Wuppertal},
             country={Germany}}

\begin{abstract}
Measurements of the inclusive and differential top-quark pair ($t\bar{t}$) cross-sections in proton–proton collisions at $\sqrt{s} = 13\,\mathrm{TeV}$, using $140\,\mathrm{fb}^{-1}$ of data collected by the ATLAS experiment at the Large Hadron Collider, are presented. Events with an opposite-charge $e\mu$ pair and $b$-tagged jets are selected, and the inclusive cross-section measurement is used to determine the top-quark pole mass via the dependence of the predicted cross-section on $m_t^\mathrm{pole}$. Complementary measurements of $e\mu b\bar{b}$ production, treating both $t\bar{t}$ and $tW$ events as signal, are also provided.
\end{abstract}




\end{frontmatter}



\section{Introduction}
\label{intro}
The study of top quark-antiquark ($t\bar{t}$) pair-production allows quantum chromodynamics (QCD) to be probed at some of the highest accessible energy scales. The large mass of the top quark, close to the scale of electroweak symmetry breaking, gives it a unique role in the Standard Model (SM) of particle physics. In addition $t\bar{t}$ production is also a significant background in many searches for physics beyond the SM. In the SM top-quark decays in almost 100\% of the cases into a $W$ boson and a
$b$-quark. In this analysis only the $t\bar{t}\rightarrow W^{+}bW^{-}\bar{b}\rightarrow e^{+}\mu^{-}\nu\bar{\nu}b\bar{b}$ 
decay channel is considered. Previous measurements of the same decay signature have been performed with increasingly precision of the inclusive cross-section $\sigma_{t\bar{t}}$. This report presents a further improved measurement of $\sigma_{t\bar{t}}$ using the full Run 2 dataset~\cite{TOPQ-2024-12} collected by the ATLAS experiment~\cite{atlas} at the Large Hadron Collider, profiting from precise calibration of lepton efficiencies, together with
recent improvements in the modelling of lepton
kinematics in $t\bar{t}$ events using the new POWHEG-BOX MiNNLO event generator~\cite{minnlops}. 
\section{Analysis strategy}
\begin{figure}[tp]
	\parbox{75mm}{\includegraphics[width=70mm]{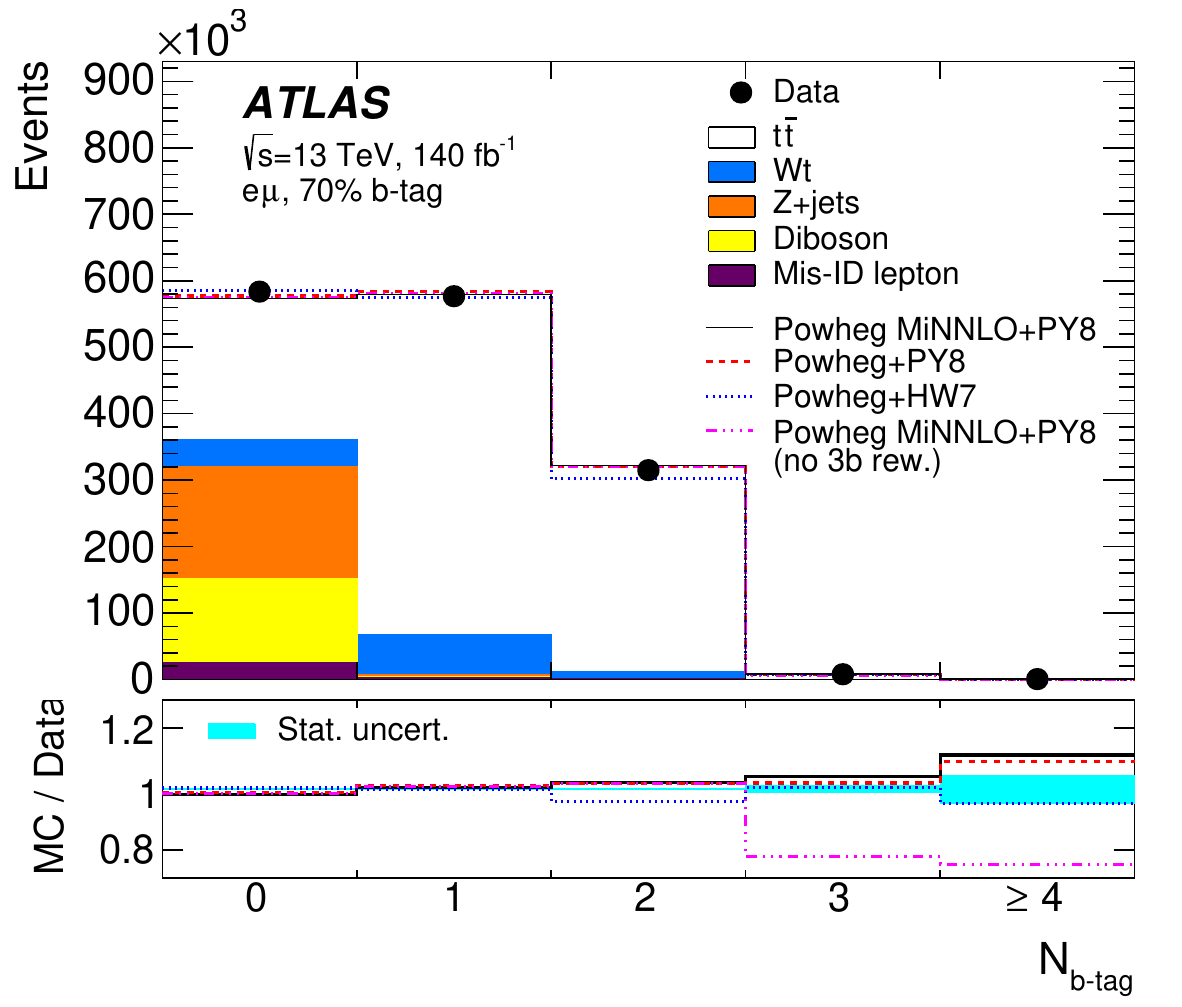}\vspace{-7mm}\center{(a)}}
	\parbox{75mm}{\includegraphics[width=70mm]{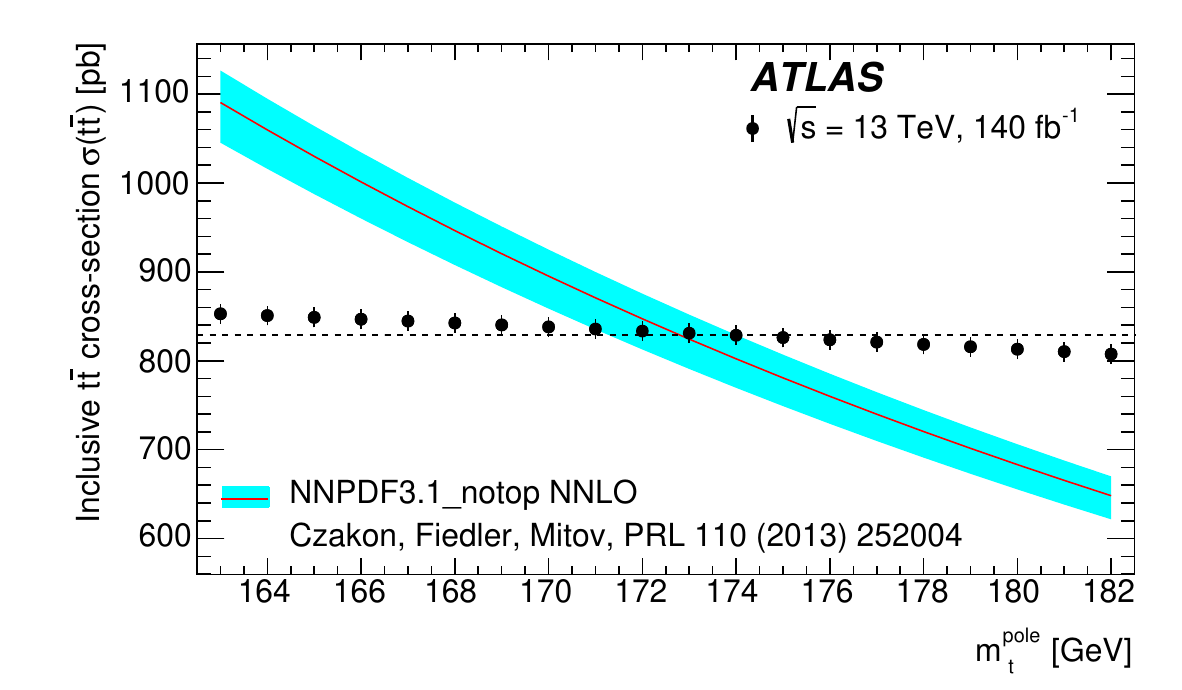}\vspace{-7mm}\center{(b)}}
	\caption{\label{f:dmcjlept}(a) Distributions of the number of $b$-tagged jets in selected opposite-charge $e\mu$ events. (b) The red line shows the predicted inclusive $t\bar{t}$ cross-section as a function of $m_t^{\mathrm{pole}}$ for the NNPDF3.1\_notop PDF set, with the cyan band indicating the total uncertainty in the prediction from PDF+$\alpha_s$ and QCD scale uncertainties. The black points show the experimental
		measurement with its uncertainty and dependence on the assumed value of
		$m_t$ through acceptance and background corrections. The dotted line shows the experimental central value at a fixed top quark mass of 172.5\,GeV~\cite{TOPQ-2024-12}.}
\end{figure}

This analysis makes use of reconstructed electrons, muons and $b$-tagged jets.
Jets were reconstructed using the anti-$k_t$ algorithm
with radius parameter $R=0.4$. Jets likely to contain $b$-hadrons are tagged using a multivariate discriminant based on deep-learning techniques making use of track impact parameters and reconstructed secondary vertices.
Events with an opposite-charge $e\mu$ pair comprise the main
analysis sample, whilst events with a same-charge $e\mu$ pair were used to
estimate the background from misidentified leptons.
The inclusive $\sigma_{t\bar{t}}$ and the $t\bar{t}\rightarrow e\mu$ differential cross-sections are measured
using a double-tagging technique employing subsets of the opposite-charge $e\mu$ sample with exactly one and exactly two $b$-tagged jets.
The distribution of the $b$-tagged jet multiplicity $n_\mathrm{b-tag}$ is shown in Figure~\ref{f:dmcjlept}(a) and compared with the
expectation from simulation.
The numbers of opposite-charge $e\mu$ events with exactly one ($N_1$) and exactly two ($N_2$) $b$-tagged jets are then counted.
The two event counts satisfy the tagging equations:
\begin{equation}
	\begin{array}{lll}
		N_1 & = & L \sigma_{t\bar{t}} \epsilon_{e\mu} 2\epsilon_b (1-C_b \epsilon_b) + N_1^{\mathrm{bkg}} , \\*[2mm]
		N_2 & = & L \sigma_{t\bar{t}} \epsilon_{e\mu} C_b \epsilon_b^2 + N_2^{\mathrm{bkg}},
	\end{array}\label{e:tags}
\end{equation}
where $L$ is the integrated luminosity, $\epsilon_{e\mu}$ the efficiency for a $t\bar{t}$ event to pass the opposite-charge $e\mu$
selection, and $C_b$ is a tagging correlation coefficient close to unity.
The combined probability for a jet from the quark $q$ in the $t\rightarrow Wq$
decay to fall within the acceptance of the detector, be reconstructed as a high-$p_T$ jet, and be tagged as a $b$-jet, is denoted by $\epsilon_b$.
The number of background events $N_1^{\mathrm{bkg}}$ and $N_2^{\mathrm{bkg}}$  are evaluated using a combination of simulation
and data control samples. The tagging equations are then solved to determine $\sigma_{t\bar{t}}$ and $C_b$.
The fiducial cross-section $\sigma_{t\bar{t}}^\mathrm{fid}$ corresponding to the production of $t\bar{t}$ events with a particle-level electron and muon from $W \rightarrow e/\mu$ decay satisfying transverse momentum $p_\mathrm{T} > 20$ GeV and pseudorapidity $|\eta| < 2.5$ is $\sigma_{t\bar{t}}^\mathrm{fid}=A_{e\mu}\sigma_{t\bar{t}}$, and can be measured by replacing $\sigma_{t\bar{t}} \epsilon_{e\mu}$ with $\sigma_{t\bar{t}} G_{e\mu}$ in Equations~(\ref{e:tags}), where $A_{e\mu}$ corresponds to the fiducial acceptance and $G_{e\mu}$ to the reconstruction efficiency. 

\section{Results}
The inclusive cross-sections are determined separately for the different data-taking periods in Run 2 and combined using BLUE in order to benefit from only partially correlated uncertainties of the luminosity measurement. The results are then:
\begin{eqnarray*}
	\sigma_{t\bar{t}}  &=&  829.3 \pm 1.3 \pm 8.0 \pm 7.3 \pm 1.9 \, \mathrm{pb} \\ 
	\sigma_{t\bar{t}}^\mathrm{fid}  &=&  14.04 \pm 0.02 \pm 0.1 \pm 0.12 \pm 0.03 \, \mathrm{pb},\\
\end{eqnarray*}
where the four uncertainties are due to data statistics, experimental and
theoretical  systematic effects, the knowledge of 
the integrated luminosity, and the knowledge of the LHC beam energy.
The result for $\sigma_{t\bar{t}}$ is reported for $m_t=172.5$\,GeV, and depends on the assumed value according to
$(1/\sigma_{t\bar{t}})\, \mathrm{d}\sigma_{t\bar{t}}/\mathrm{d}m_t=-0.29$\%/GeV.
The $m_t$ dependence of the measured $\sigma_{t\bar{t}}^\mathrm{fid}$ is negligible.

The strong dependence of the prediction for $\sigma_{t\bar{t}}$ on $m_t^{\mathrm{pole}}$ can be exploited to interpret the precise measurement of $\sigma_{t\bar{t}}$ as a measurement of $m_t^{\mathrm{pole}}$.
Figure~\ref{f:dmcjlept}(b) shows the dependence of the predicted $\sigma_{t\bar{t}}$ on $m_t^{\mathrm{pole}}$ for the NNPDF3.1\_notop PDF set,
which doesn't use top-quark data as input, calculated using {\tt Top++}~\cite{toppp}.
A Bayesian likelihood formulation was used to extract $m_t^{\mathrm{pole}}$ by maximising the compatibility with the measured $\sigma_{t\bar{t}}$,
giving a result of: $m_t^{\mathrm{pole}} = 172.8^{+1.5}_{-1.7}\,\mathrm{GeV}$.

\begin{figure}[ht]
	\parbox{83mm}{\includegraphics[width=76mm]{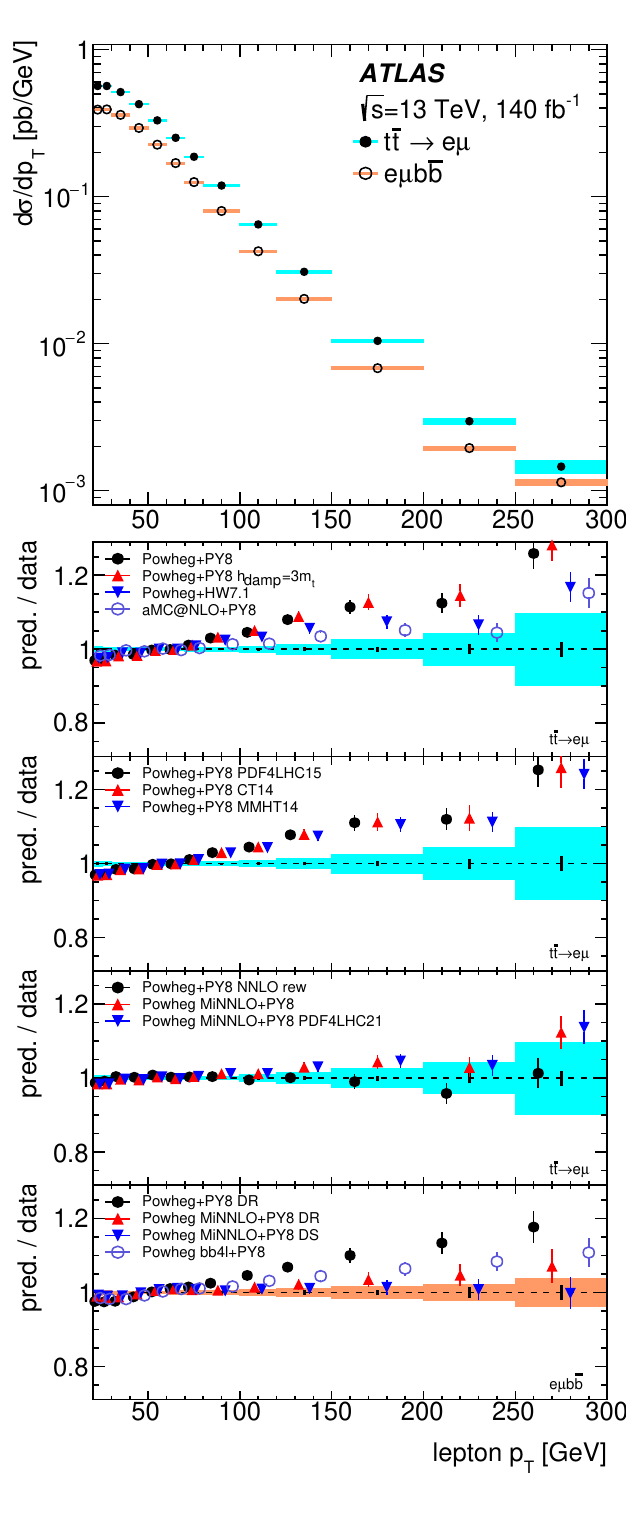}\vspace{-7mm}\center{(a)}}
	\parbox{83mm}{\includegraphics[width=76mm]{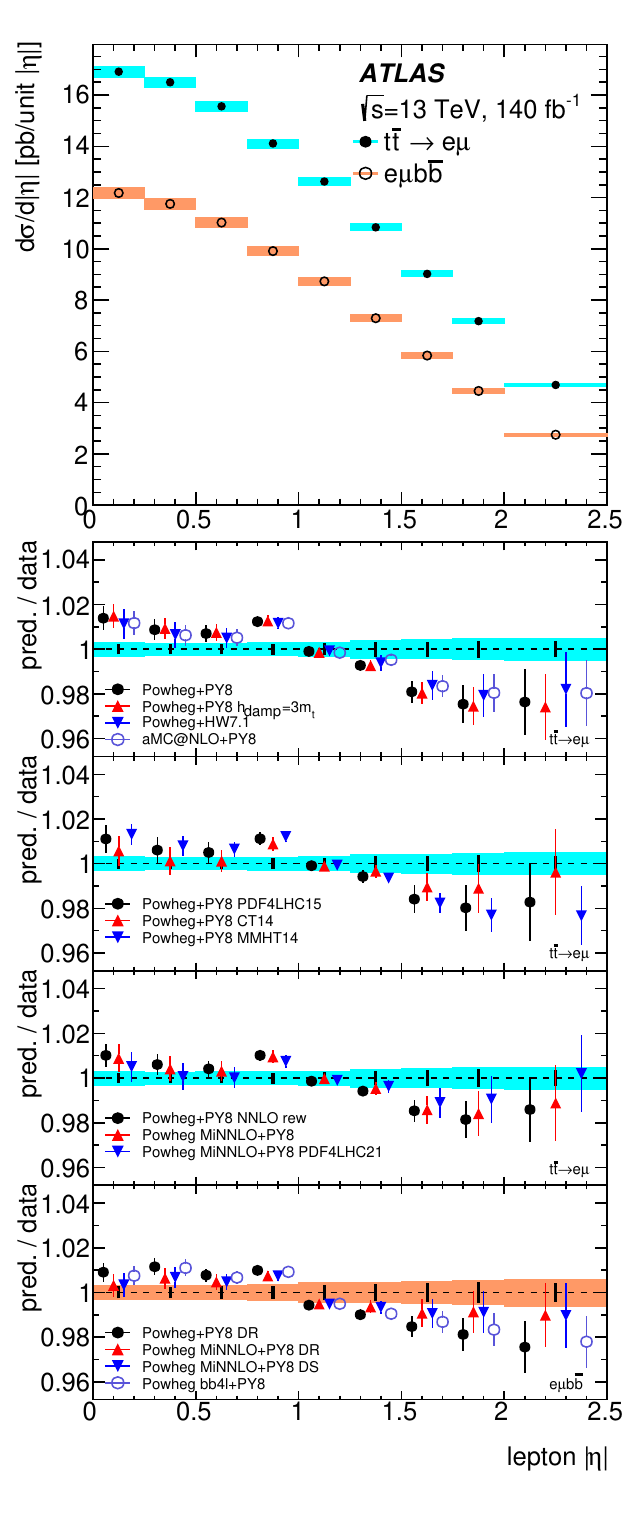}\vspace{-7mm}\center{(b)}}
	\caption{\label{f:distresa}Differential cross-sections as functions of (a) $p_\mathrm{T}$ and (b) $\eta$ of the leading lepton. The upper panels show the measured absolute $t\bar{t}$ (filled points)
		and $e\mu b\bar{b}$ (open points) cross-sections with the total uncertainties shown by the shaded bands. The other panels show ratios of various predictions to the measured normalised differential cross-sections. The markers showing the ratios for each prediction and the total
		uncertainty in the prediction is shown by the error bar. The data statistical uncertainty is shown by the black error bars and the total uncertainty by the cyan or orange band~\cite{TOPQ-2024-12}.}
\end{figure}
The same dataset is used to measure ten single-differential and
three double-differential cross-sections as functions of lepton and dilepton
kinematic variables, for both the  $t\bar{t} \rightarrow e\mu$ process and the $e\mu b\bar{b}$ final state. 
An example of a measured absolute one-dimensional $t\bar{t} \rightarrow e\mu$ and $e\mu b\bar{b}$ differential cross-sections is shown in Figure~\ref{f:distresa}. Uncertainties
as small as 0.3\% have been achieved for normalised distributions in some
parts of the fiducial regions. Comparisons with event generator predictions
show that state-of-the-art generators such as {\sc Powheg} MiNNLO or
{\sc Powheg} {\em bb4l} better model the lepton kinematics than the
{\sc Powheg} {\tt hvq} process traditionally used for LHC physics analyses.
These precise measurements provide input that can be used to further refine the modelling of top-quark production at hadron colliders.

~\\
\noindent
Copyright 2026 CERN for the benefit of the ATLAS Collaboration. Reproduction of this article or parts of it is allowed as specified in the CC-BY-4.0 license.




\end{document}